\documentclass [10pt]{article}
\usepackage {amssymb}
\usepackage {amsmath}
\usepackage [cp1251]{inputenc}
\usepackage{graphicx}
\usepackage {longtable}

\title{Singularities Motion Equations in 2-Dimensional Ideal Hydrodynamics of Incompressible Fluid}

\author{V.V.Yanovsky, A.V.Tur$^{\ast}$, K.N.Kulik}

\begin{document}

\maketitle

\textit{ Institute for Single Crystals, Nat. Academy of
Science Ukraine, Lenin Ave.60, s, Ukraine\\
E-mail address: yanovsky@isc.kharkov.ua, koskul@isc.kharkov.ua}

\textit{ $^{\ast}$Center D'etude Spatiale Des Rayonnements,
C.N.R.S.-U.P.S., 9, avenue Colonel-Roche 31028 TOULOUSE, France,
CEDEX 4. E-mail address: anatoly.tour@cesr.fr}

\textit{keyword: ideal hydrodynamic, point vortice}


\begin{abstract}
In this paper, we  have obtained motion equations for a wide class
of one-dimensional singularities in 2-D ideal hydrodynamics. The
simplest of them, are well known as point vortices. More
complicated singularities correspond to vorticity point dipoles.
It has been proved that point multipoles of a higher order
(quadrupoles and more) are not the exact solutions of
two-dimensional ideal hydrodynamics. The motion equations for a
system of interacting point vortices and point dipoles have been
obtained. It is shown that these equations are Hamiltonian ones
and have three motion integrals in involution. It means the
complete integrability of two-particle system, which has a point
vortex and a point dipole.
\end{abstract}


1. In 2-dimensional ideal hydrodynamics of incompressible fluid,
point vortices play an exceptionally important role. Actually,
point vortices are the well-defined quasiparticles. This means
that various flows of ideal fluid can be represented as movements
induced by a system of interacting point vortices \cite{ref1},
\cite{ref19}. Certain restrictions of this point of view are
discussed, in particular, in \cite{ref2}. It is well known that
the dynamics of three point vortices is regular while four or more
point vortices dynamics is chaotic \cite{ref3}, \cite{ref4},
\cite{ref8}, \cite{ref20}. The generalization of such point
vortices on sphere or in isolated areas was already done by some
authors \cite{ref5}, \cite{ref6}, (see also
\cite{ref19},\cite{ref16}). Point vortices are also used for
two-dimensional turbulence models \cite{ref7}-\cite{ref10} and for
studies of 2D turbulence spectra. The nature of vortices
interaction and features of statistically stationery states
generation are essential \cite{ref11}, \cite{ref12}, as well as
the nature of singular velocity field, induced by point vortices
movements \cite{ref13}. Besides, point vortices engender more
complicated solutions of 2D Euler equation and magnetic fields
configuration in MHD \cite{ref14}, \cite{ref14a} \cite{ref15}. In
this paper, we will show that 2D Euler equation has exact
solutions with more complicate singularities, such as a set of
point vortex dipoles. Any finite sum of sets of point vortex
dipoles and of point vortices is also an exact solution for 2D
Euler equation with the moving singularities. Motion equations for
a system of interacting vortices and point dipoles have been
obtained. It is shown that these equations are Hamiltonian ones
and have three motion integrals in involution.

2. In this part, we are going to obtain motion equations of
different kind of one-dimensional singularity for two-dimensional
Euler equations. Let us start with two-dimensional case of Euler
equation for incompressible fluid:
\begin{equation}\label{eq1}
    \frac{\partial V_i}{\partial t}+V_j \frac{\partial V_i}{\partial
    x_j}=-\frac{\partial P}{\partial x_i}
\end{equation}

\[div \vec{V}=0\]
Now we shall use the potential $\varphi$ of velocity field
$\vec{V}$, which is defined as:
\begin{equation}\label{eq2}
     V_i =\varepsilon_{ik}\frac{\partial \varphi}{\partial x_k}
\end{equation}
Here $i=1,2$ and $\varepsilon_{ik}$ is a unit antisymmetric
tensor. After excluding pressure we shall consider a well known
form of Euler equation:

\begin{equation}\label{eq3}
    \frac{\partial \Delta \varphi}{\partial t}+\{\Delta
    \varphi,\varphi\}=0
\end{equation}
Here $\Delta$ is two-dimensional Laplace operator, and
$\{A,B\}=\varepsilon \frac{\partial A}{\partial x_i}
\frac{\partial B}{\partial x_k}$ is Poisson's bracket. From
physical point of view, this means the freezing of vorticity field
$\omega$ into the fluid. Now the vorticity $\omega$ can be defined
as:

\begin{equation}\label{eq4}
    \omega=-\Delta \varphi
\end{equation}
Let us suppose, that vorticity field singularities can be defined
in terms of generalized functions as:
\[ -\Delta
    \varphi=\sum_{\alpha=1}^{N}\Gamma_{\alpha}\delta(\vec{x}-\vec{x}_{v}^{\alpha}(t))
    +\]
\[+\sum_{\beta=1}^{M}D_{m}^{\beta}(t)
    \frac{\partial \delta(\vec{x}-\vec{x}_{d}^{\beta}(t))}{\partial
    x_m}+\]
\begin{equation}\label{eq5}
      +\sum_{\gamma=1}^{K}\mu_{i_1 i_2}^{\gamma}(t) \frac{\partial^2 \delta(\vec{x}-\vec{x}_{\mu}^{\gamma}(t))}{\partial
    x_{i_1} \partial x_{i_2}} + \cdots
\end{equation}
In other words, vorticity field is represented as a sum of point
multipoles and point vortices. Greek indexes in the right part
numerate the corresponding objects. The first sum corresponds to
point vortices; $\Gamma_\alpha$ is vortex stretch and
$\vec{x}_{v}^{\alpha}(t)$ is the coordinate of $\alpha$-th vortex.
The second sum corresponds to point dipoles; $D_{m}^{\beta}(t)$ is
dipole moment and  $\vec{x}_{d}^{\beta}(t)$ is the coordinate of
$\beta$-th point dipole. The next contributions correspond to
multipoles of a higher order, e.g. $\mu_{i_1 i_2}^{\gamma}(t)$ is
quadrupole moment (symmetric traceless tensor) and
$\vec{x}_{\mu}^{\gamma}(t)$ is the coordinate of $\gamma$-th point
quadrupole. The sense of the designation is the same as described
above.

It is easy to obtain the explicit form of the potential from
equation (\ref{eq5})
\[\varphi=-\frac{1}{4 \pi}\sum_{\alpha=1}^{N}\Gamma_{\alpha}\ln|\vec{x}-\]
\[-\vec{x}_{v}^{\alpha}(t)|-\frac{1}{2 \pi}\sum_{\beta=1}^{M}D_{l}^{\beta}(t)
    \frac{(x_l-x_{ld}^{\beta}(t))}{|\vec{x}-\vec{x}_{d}^{\,\beta}(t)|^{\,2}}+\]

\begin{equation}\label{eq6}
    -\frac{1}{4 \pi} \sum_{ \gamma =1}^{N} \mu_{i_1 , i_2}^{\gamma}
    (t) \frac{\partial^2 \ln |\vec{x}-\vec{x}^{\gamma}_{\mu}(t)|}{\partial x_{i_1} \partial x_{i_2}} \cdots
\end{equation}
The main problem is to find under which conditions the equation
(\ref{eq6}) will be compatible with Euler equation (\ref{eq3}). In
order to solve this problem, we have to substitute the development
(\ref{eq6}) directly into Euler equation (\ref{eq3}). Such
substitution of the equations (\ref{eq5}) and (\ref{eq6}) into
equation (\ref{eq3}) leads to the following:
\[\sum_{\alpha=1}^{N}\Gamma_{\alpha}\left(\frac{d x_{vi}^{\alpha}(t)}{d t}- V_i
    \right)\frac{\partial \delta (\vec{x}-\vec{x}_{v}^{\alpha}(t)) }{\partial
    x_i}-\]
\[ -\sum_{\beta=1}^{M}\frac{d D_{i}^{\beta}(t)}{d t} \cdot \frac{\partial \delta (\vec{x}-\vec{x}_{d}^{\beta}(t)) }{\partial
    x_i}-\]
\[-\sum_{\beta=1}^{M}D_{m}^{\beta}(t)\left(\frac{d x_{di}^{\beta}(t)}{d t}- V_i
    \right)\frac{\partial^2 \delta (\vec{x}-\vec{x}_{d}^{\beta}(t))}{\partial
    x_m\partial x_i}-\]
\[- \sum_{ \gamma =1}^{N} \frac{d {\mu}_{i_1 , i_2}^{\gamma}
    (t)}{d t} \cdot \frac{\partial^2 \delta (\vec{x}-\vec{x}^{\gamma}_{\mu}(t))}{\partial x_{i_1} \partial
    x_{i_2}}-\]

\begin{equation}\label{eq7}
    - \sum_{ \gamma =1}^{N} \mu_{i_1 , i_2}^{\gamma}
    (t) \left( \frac{d {x}_{\mu i}^{\gamma}(t)}{dt} -V_{i}\right)\frac{\partial^3 \delta (\vec{x}-\vec{x}_{\mu}^{\gamma} (t))}{\partial x_{i_1} \partial x_{i_2} \partial
    x_{i}}+ \cdots=0
\end{equation}
Here $V_i$ means the components of velocity field calculated with
equations  (\ref{eq6}) and (\ref{eq2}). To make it less bulky, we
shall write down the dipole and quadrupole contributions only.
There is no special difficulties in taking into account the
contributions of higher order multipoles. Let us note that
$\delta$- functions in the equation (\ref{eq7}) are the functions
of both $t$, and  $\vec{x}$. Now we shall deal with equation
(\ref{eq7}) using properties of the generalized functions. In
particular, the following well known properties of  $\delta$ -
functions derivatives will be very important (see, for example,
\cite{ref17},\cite{ref18}):
\[\alpha(\vec{x})\frac{\partial \delta (\vec{x}-\vec{A})}{\partial
    x_i}=\alpha(\vec{x})|_{\vec{x}=\vec{A}}  \frac{\partial \delta (\vec{x}-\vec{A})}{\partial
    x_i}-\]
\begin{equation}\label{eq8}
    - \left.\frac{\partial \alpha (\vec{x})}{\partial
    x_i} \right|_{\vec{x}=\vec{A}}  \delta(\vec{x}-\vec{A})
\end{equation}

and
\[\alpha(\vec{x})\frac{\partial^2 \delta (\vec{x}-\vec{A})}{\partial
    x_i \partial x_j}=\alpha(\vec{x})|_{\vec{x}=\vec{A}}  \frac{\partial^2
    \delta (\vec{x}-\vec{A})}{\partial x_i \partial
    x_j}-\]
\[-\left.\frac{\partial \alpha (\vec{x})}{\partial x_i} \right|_{\vec{x}=\vec{A}} \frac{\partial
\delta(\vec{x}-\vec{A})}{\partial x_j}-\]
\begin{equation}\label{eq9}
     -\left.\frac{\partial \alpha
(\vec{x})}{\partial x_j} \right|_{\vec{x}=\vec{A}} \frac{\partial
\delta(\vec{x}-\vec{A})}{\partial x_i}+\left.\frac{\partial^2
\alpha(\vec{x})}{\partial x_i \partial
x_j}\right|_{\vec{x}=\vec{A}} \delta(\vec{x}-\vec{A})
\end{equation}
 \[\ldots\]
When using these relations, we can obtain  $\delta$ - functions
derivatives coefficients, depending only on time. Then, the
equation (\ref{eq7}) takes the form:
\[\sum_{\alpha=1}^{N}\Gamma_{\alpha}\left(\frac{d x_{vi}^{\alpha}(t)}{d t}-
\left.V_i\right|_{\vec{x}=\vec{x}_{v}^{\alpha}(t)}
    \right)\frac{\partial \delta (\vec{x}-\vec{x}_{v}^{\alpha}(t)) }{\partial
    x_i}-\]
\[-2 \sum_{\beta=1}^{M}\left(\frac{d D_{i}^{\beta}(t)}{d t}-\left.D_{m}^{\beta}(t)
    \frac{\partial V_i}{\partial x_m}\right|_{\vec{x}=\vec{x}_{d}^{\beta}(t)}
    \right)\frac{\partial \delta (\vec{x}-\vec{x}_{d}^{\beta}(t)) }{\partial
    x_i}+\]

\[+2 \sum_{\beta=1}^{M}D_{m}^{\beta}(t)\left(\frac{d x_{di}^{\beta}(t)}{d t}-
\left.V_i\right|_{\vec{x}=\vec{x}_{d}^{\beta}(t)}
    \right)\frac{\partial^2 \delta (\vec{x}-\vec{x}_{d}^{\beta}(t)) }{\partial
    x_m \partial x_i}+\]

\[+\sum_{ \gamma =1}^{N} \left( \frac{d {\mu}_{i_1 , i_2}^{\gamma}
    (t)}{d t} - \mu_{i_3 , i_2}^{\gamma}(t) \left(\left. \frac{\partial
 V_{i_1}(\vec{x})}{
\partial x_{i_3} }\right|_{\vec{x}=\vec{x}_{\mu}^{\gamma}(t)} \right) - \right. \]

\[\left. -\mu_{i_1 , i_3}^{\gamma}(t) \left(\left. \frac{\partial
 V_{i_2}(\vec{x})}{
\partial x_{i_3} }\right|_{\vec{x}=\vec{x}_{\mu}^{\gamma}(t)} \right)\right)\cdot \frac{\partial^2  \delta (\vec{x}-\vec{x}_{\mu}^{\gamma}(t))}{\partial x_{i_1} \partial
    x_{i_2}}-\]
\[-\sum_{ \gamma =1}^{N} \mu_{i_1 , i_2}^{\gamma}
    (t) \left( \frac{d {x}_{\mu i_3}^{\gamma}(t)}{d t} -  \left( \left.
V_{i_3}(\vec{x})\right|_{\vec{x}=\vec{x}_{\mu}^{\gamma}(t)}
\right) \right) \cdot\]
\[\cdot \frac{\partial^3 \delta
(\vec{x}-\vec{x}_{\mu}^{\gamma}(t))}{\partial x_{i_1} \partial
x_{i_2}
\partial
    x_{i_3}}+\]
\[+\sum_{ \gamma =1}^{N} \mu_{i_1 ,
i_2}^{\gamma}(t) \left(\left. \frac{\partial^2 V_{i_3}(\vec{x})}{
\partial x_{i_1} \partial
x_{i_2}}\right|_{\vec{x}=\vec{x}_{\mu}^{\gamma}(t)} \right)\cdot\]
\begin{equation}\label{eq10}
     \cdot
\frac{\partial  \delta
(\vec{x}-\vec{x}_{\mu}^{\gamma}(t))}{\partial
 x_{i_3}}+\cdots=0
 \end{equation}
From (\ref{eq10}) we can obtain the motion equation for
interacting singularities and the evolution equations for
multipole moments. In order to do it, we need  to have all the
coefficients equal to zero before different  $\delta$ - functions
derivatives independently. If all these equations are compatible,
then the velocity field, generated by potential (\ref{eq6}), is
the exact generalized solution of two-dimensional Euler equation.
It is easy to see that if coefficients before generalized
functions relative to vortexes and dipoles tend to zero, then it
gives vortex and dipoles motion equations only, as well as
evolution law for dipole moment. Further, it will be shown that
equations of this set are compatible. But the situation is getting
substantially different starting from the quadrupole vortices. As
a matter of fact, higher order multipole moments give a set of
uncompatible equations. In general case, the compatibility
conditions are satisfied only, if higher order multipole moments
starting from quadrupole one, are zero. Hence, in two-dimensional
hydrodynamics only sets of point vortices and point dipoles give
generalized point solutions. Let us note that singularities motion
equations have natural physical meaning of the vorticity freezing
into medium motions. Thus, the motion velocity of a chosen
singularity coincides with the medium velocity in the same point,
induced by all other singularities. From mathematical point of
view, the selfinteraction is absent in motion equations for dipole
singularities as well as for point vortices. Unlike point vortices
when vortex stretch is constant, the dipole moment is function of
time. Let us give now the final equations for evolution of
interacting point vortices and point dipoles:
\begin{equation}\label{eq11}
    \frac{d \vec{x}_{vi}^{\alpha}(t)}{d t}=-\varepsilon_{ik} \left\{\sum_{\gamma \neq \alpha}^{N}
      \frac{\Gamma_{\gamma}}{2 \pi}
    \frac{(x_{vk}^{\alpha}-x_{vk}^{\gamma})}{|\vec{x}_{v}^{\alpha}-\vec{x}_{v}^{\gamma}|^2}\right.+
\end{equation}
\[+\sum_{\beta}^{M} \frac{D_{l}^{\beta}(t)}{\pi}\left( \left. \frac{\delta_{lk}}{|\vec{x}_{v}^{\alpha}-
\vec{x}_{d}^{\beta}|^2}-\frac{2(x_{vl}^{\alpha}-x_{dl}^{\beta})(x_{vk}^{\alpha}-x_{dk}^{\beta})}
{|\vec{x}_{v}^{\alpha}-\vec{x}_{d}^{\beta}|^4} \right)\right\}=0\]

\begin{equation}\label{eq12}
    \frac{d \vec{x}_{di}^{\beta}(t)}{d t}=-\varepsilon_{ik} \left\{\sum_{\alpha}^{N}
      \frac{\Gamma_{\alpha}}{2\pi}
    \frac{(x_{dk}^{\beta}-x_{vk}^{\alpha})}{|\vec{x}_{v}^{\alpha}-\vec{x}_{d}^{\beta}|^2}\right.+
\end{equation}
\[+\sum_{\gamma \neq \beta}^{M} \frac{D_{l}^{\gamma}(t)}{\pi}\left( \left. \frac{\delta_{lk}}{|\vec{x}_{d}^{\beta}-
\vec{x}_{d}^{\gamma}|^2}-\frac{2(x_{dl}^{\beta}-x_{dl}^{\gamma})(x_{dk}^{\beta}-x_{dk}^{\gamma})}
{|\vec{x}_{d}^{\beta}-\vec{x}_{d}^{\gamma}|^4} \right)\right\}=0\]
\[\frac{d D_{i}^{\beta}(t)}{d t}=D_{m}^{\beta}(t)\varepsilon_{ik} \times\]
\[\times \left\{\sum_{\alpha}^{N}\frac{\Gamma_{\alpha}}{2\pi}\left(
 \frac{\delta_{km}}{|\vec{x}_{d}^{\beta}-\vec{x}_{v}^{\alpha}|^2}-\frac{2(x_{dk}^{\beta}-x_{vk}^{\alpha})(x_{dm}^{\beta}-x_{vm}^{\alpha})}
{|\vec{x}_{d}^{\beta}-\vec{x}_{v}^{\alpha}|^4} \right)+ \right.\]

\[+\sum_{\gamma \neq \beta}^{M} \frac{D_{l}^{\gamma}(t)}{\pi}\left(-
\frac{\delta_{lk}2(x_{dm}^{\beta}-x_{dm}^{\gamma})}{|\vec{x}_{d}^{\beta}-
\vec{x}_{d}^{\gamma}|^4}-\frac{\delta_{ml}2(x_{dk}^{\beta}-x_{dk}^{\gamma})}
{|\vec{x}_{d}^{\beta}-\vec{x}_{d}^{\gamma}|^4} -\right.\]
\[-\frac{\delta_{mk}2(x_{dl}^{\beta}-x_{dl}^{\gamma})}
{|\vec{x}_{d}^{\beta}-\vec{x}_{d}^{\gamma}|^4} +\]
\begin{equation}\label{eq13}
    \left.\left.+\frac{8(x_{dl}^{\beta}-x_{dl}^{\gamma})(x_{dk}^{\beta}-x_{dk}^{\gamma})(x_{dm}^{\beta}-x_{dm}^{\gamma})}
{|\vec{x}_{d}^{\beta}-\vec{x}_{d}^{\gamma}|^6}\right) \right\}=0
\end{equation}
Equations (\ref{eq11}) for zero values of dipole moment coincide
exactly with equations for point vortices \cite{ref1}. Equations
(\ref{eq11})-(\ref{eq13}) give a set of compatible equations,
describing the evolution of interacting  $N$ - point vortices and
$M$ - point dipoles. The set of equations
((\ref{eq12})-(\ref{eq13}) for $\Gamma_\alpha \equiv 0$, describes
the evolution of $M$ - points dipoles.

3. Let us consider now some general properties of equations
(\ref{eq11}-\ref{eq13}). First of all this system is Hamiltonian
with the following Hamiltonian:
\[H=-\sum_{\alpha \neq \beta}^{N}\frac{\Gamma_\alpha \Gamma_\beta}{4 \pi}
    \ln{|\vec{x}_{v}^{\alpha}-\vec{x}_{v}^{\beta}|}-\sum_{\alpha,\beta}^{N,M}
    \frac{\Gamma_\alpha}{2 \pi}\frac{ D_{\beta l}(x_{vl}^{\alpha}-x_{dl}^{\beta})}
    {|\vec{x}_{v}^{\alpha}-\vec{x}_{d}^{\beta}|^2}-\]
\[-\sum_{\substack{\beta < \gamma=1\\\beta \neq \gamma}}^{M} \frac{2D_{\beta m}}{\pi}\frac{
(x_{dm}^{\beta}-x_{dm}^{\gamma})D_{\gamma l}
(x_{dl}^{\beta}-x_{dl}^{\gamma})}
 {|\vec{x}_{d}^{\beta}-\vec{x}_{d}^{\gamma}|^4}+\]
\begin{equation}\label{eq14}
 +\sum_{\substack{\beta < \gamma=1\\\beta \neq \gamma}}^{M} \frac{D_{\beta m}D_{\gamma
 m}(\vec{x}_{d}^{\beta}-\vec{x}_{d}^{\gamma})^2}{|\vec{x}_{d}^{\beta}-\vec{x}_{d}^{\gamma}|^4}
\end{equation}
First group of terms is a usual interaction energy of point
vortices, the second group is  vortexes and dipoles interaction
energy, and the third one is dipoles-dipoles interaction. Motion
equations (\ref{eq11}-\ref{eq13}) have the following Hamiltonian
form:
 \begin{equation}\label{eq15}
    \Gamma_{\alpha} \dot{x}_{v
    i}^{\alpha}=\varepsilon_{ik}\frac{\partial H}{\partial x_{v
    k}^{\alpha}}
\end{equation}

\begin{equation}\label{eq16}
     \dot{x}_{d i}^{\beta}=-\varepsilon_{ik}\frac{\partial H}{\partial D_{k}^{\beta}}
\end{equation}

\begin{equation}\label{eq17}
    \dot{D}_{i}^{\beta}=-\varepsilon_{ik}\frac{\partial H}{\partial
    x_{d k}^{\alpha}}
\end{equation}
It is well known that for vortices, conjugated canonical variables
are its coordinates. For point dipoles, conjugated canonical
variables are dipole moments and dipole coordinates. The set of
equations (\ref{eq15}-\ref{eq17}) has some integrals of motion,
which are easy to check by direct calculation:
 \begin{equation}\label{eq18}
    I_1=\sum_{\alpha=1}^{N}\Gamma_\alpha
    x_{v1}^{\alpha}-\sum_{\beta=1}^{M} D_{1}^{\beta}=const
\end{equation}

\begin{equation}\label{eq19}
    I_2=\sum_{\alpha=1}^{N}\Gamma_\alpha
    x_{v2}^{\alpha}-\sum_{\beta=1}^{M} D_{2}^{\beta}=const
\end{equation}

\begin{equation}\label{eq20}
    J=\sum_{\alpha=1}^{N}\Gamma_\alpha
    (\vec{x}_{v}^{\alpha})^2-2\sum_{\beta=1}^{M} (\vec{D}^{\beta} \cdot \vec{x}_{d}^{\beta})=const
\end{equation}
These conservation laws are, in a sense, quite similar to the
conservation laws, known for point vortices systems. Of course,
the energy of point vortices and point dipoles system is
conserved, so $H=const$.

The set of equations (\ref{eq15}-\ref{eq17}) can be written in a
more symmetric way with the Poisson's bracket:
\[\{A,B\}=\sum_{\alpha} \frac{1}{\Gamma_{\alpha}} \varepsilon_{ik}
    \frac{\partial A}{\partial x_{i}^{\alpha}}\frac{\partial B}{\partial x_{k}^{\alpha}}
    -\]
 \begin{equation}\label{eq21}
    -\sum_{\beta}\varepsilon_{ik}
    \frac{\partial A}{\partial D_{i}^{\beta}}\frac{\partial B}{\partial x_{dk}^{\beta}}
    -\sum_{\beta}\varepsilon_{ik}
    \frac{\partial A}{\partial x_{di}^{\beta}}\frac{\partial B}{\partial D_{k}^{\beta}}
\end{equation}
Then the Hamilton equations take the canonical form:
\begin{equation}\label{eq22}
    \dot{x}_{vi}^{\alpha}=\{x_{vi}^{\alpha},H\}
\end{equation}
\begin{equation}\label{eq23}
    \dot{x}_{di}^{\beta}=\{x_{di}^{\beta},H\}
\end{equation}
\begin{equation}\label{eq24}
    \dot{D}_{i}^{\beta}=\{D_{i}^{\beta},H\}
\end{equation}
Using Poisson's bracket (\ref{eq21}), we can check, that only
three from the above mentioned invariants are in involution.

Actually, for four above mentioned motion integrals, one can get:
\begin{equation}\label{eq25}
   \{I_1,H\}=\{I_2,H\}=\{J,H\}
\end{equation}
besides, we obtain :
\begin{equation}\label{eq26}
    \{I_1,I_2\}=\sum_{\alpha=1}^{N}\Gamma_\alpha
\end{equation}

\begin{equation}\label{eq27}
    \{I_1,J\}=2I_2
\end{equation}

\begin{equation}\label{eq28}
    \{I_2,J\}=-2I_1
\end{equation}
In other words, these integrals are not in involution. We can find
by the use of Poisson's bracket that
\begin{equation}\label{eq29}
   \{I_{1}^{2}+I_{2}^{2},J\}=\{I_{1}^{2},J\}+\{I_{2}^{2},J\}=
\end{equation}
\[=2I_1\{I_1,J\}+2I_2\{I_2,J\}=2I_1 2I_2-2I_2 2I_1=0\]
Then, the first three motion integrals are in involution
$H,J,I_{1}^{2}+I_{2}^{2}$. A particular case of these conservation
laws for a system of vortices is well-known
\cite{ref1},\cite{ref5}. According to Liouville's theorem, this
means that one point vortex interacting with one point dipole
constitute a completely integrable system. Let us remind that for
the same reason, the system of three point vortices is also
completely integrable \cite{ref19, ref20}.


\begin{thebibliography}{99}


\bibitem{ref1}H.Lamb, Hydrodynamics, Ed. 6-th., N. Y. Dover publ. 1945.

\bibitem{ref19}P.G.Saffinan, Vortex Dynamics, Camb. Univ. Press., 1992.

\bibitem{ref2}H.M.Glaz, Two attempts at modeling two-dimensional turbulence, in <<Turbulence Seminae>>,
ed.P.Bernard, T.Ratio, Lecture Notes in Mathematics,
Springer-Verlag, Berlin-Heidelberg-New York, 1977.

\bibitem{ref3}E.A.Novikov, Y.B.Sedov, Stochastic properties of a four-vortex system. Sov. Phys. JETP, 48(3), p.440-444,
1978.

\bibitem{ref4}S.L.Ziglin, Non-integrability of the problem of motion of four point vortices,
Doklady AN SSSR, v. 250, No.6, p. 1296–1300,  1979. (in russin)

\bibitem{ref8}H.Aref, Integrable, chaotic and turbulent vortex motion in two-dimensional
flows, Ann. Rev. Fluid Mech., 15, 345-89, 1983.

\bibitem{ref20}H.Aref, Motion of three vortices, Phys. Fluids, v. 31,
No.6, p. 1392-1409, 1988.


\bibitem{ref5}P.K.Newton, The N-Vortex problem. Analytical Techniques,
Springer, 2001.

\bibitem{ref6}Y.Kimura, Motion of point vorteces in a circular
domain, J.Phys.Soc.Japan, 57, No.5, p.1641-1649, 1988.

\bibitem{ref16}V.V.Meleshko, M.Yu.Konstantinov, Vortex Dynamics and Chaotic Phenomena,
World Scientific, Singapore, 1999.

\bibitem{ref7}W.D.McComb, The Physics of Fluid
Turbulence, Clarendon Press, Oxford, 1992.

\bibitem{ref9}R.H.Kraichnan, Statistical dynamics of two-dimensional flows, J. Fluid
Mech. 67, 155-175, 1975.

\bibitem{ref10}Y.B.Pointin,  T.S.Lundgren,  Statistical mechanics of two-dimensional
vortices in a bounded container, Phys. Fluids, 109, p.1459-70,
1976.

\bibitem{ref11}K.Fine, A.Cass,  W.Flynn,  C.Dryscoll, Relaxation of 2D
Turbulence to Vortex Crystal. Phys. Rev. Lett. 1995, 75, p. 3277.

\bibitem{ref12}L.Onsager, Statistical Hydromechanics, Nuovo Cimento
Suppl. Al., v.VI, ser.IX, 1949.

\bibitem{ref13}E.A.Novikov, Dynamics and statistics of a system of vortices, Sow Phys.
JETP. 41, 937-43, 1976.

\bibitem{ref14}V.V.Yanovsky, A.V.Tur, P.Louarn, D.Le Queau,  On the link between the two-dimensional
hydrodynamics and the three-dimensional (3-D) magnetostatic: A new
method for obtaining a 3-D solution of the magnetostatic
equlibrium, Physics of plazmas, No.9, v.8, p.4255-4258, 2001.

\bibitem{ref14a}D.Crowdy, A class of exact multipolar vortices, Phys.Fluid, v.11, No.9,
p.2556-2564, 1999.

\bibitem{ref15}A.V.Tur, V.V.Yanovsky, Point vortices with a rational necklace: New exact stationary
solutions of the two-dimensional Euler equation, Physycs of
Fluids, Vol.16, No.8, p.2877-2885, 2004.

\bibitem{ref17}I. M.Gelfand, G.E.Shilov, Generalized functions, Academic Press, New York and London,
1967.

\bibitem{ref18}V.S.Vladimirov, Generalized functions in mathematical physics Moscow, Nauka, 1979.





\end{thebibliography}
\end{document}